\documentclass[12pt,a4paper]{article}
\usepackage{epsfig}
\begin{document}
\textwidth=135mm
 \textheight=200mm
 
\begin{center}
{\bfseries Mott-hadron resonance gas and lattice QCD thermodynamics
}
\vskip 5mm
D.~Blaschke$^{\star,\dag}$\footnote{Email: blaschke@ift.uni.wroc.pl},
A.~Dubinin$^{\star}$, L.~Turko$^{\star}$
\vskip 5mm
{\small {\it $~^{\star}$ Instytut Fizyki Teoretycznej, Uniwersytet Wroc{\l}awski, 50-204 Wroc{\l}aw, Poland }}\\
{\small {\it $^\dag$ BLTP, Joint Institute for Nuclear Research, 141980 Dubna, Russia}} \\
\end{center}

\centerline{\bf Abstract}
We present an effective model for the generic behaviour of hadron masses and phase shifts at finite
temperature which shares basic features with recent developments within the PNJL model for correlations in quark matter.
On this basis we obtain the transition between a hadron resonance gas phase and the quark gluon plasma in the spirit of the generalized Beth-Uhlenbeck approach where the Mott dissociation of hadrons is encoded in the hadronic phase shifts.
Here we restrict ourselves to low-lying hadronic channels and perform a discussion of recent lattice QCD thermodynamics results from this perspective.
We find agreement in the asymptotic regions while for the description of the transition  itself the inclusion of further hadronic channels as well as a selfconsistent determination of the continuum thresholds is required.
\vskip 10mm
\section{\label{sec:intro}Introduction}
Simulations of lattice QCD (LQCD) are in practice the only reliable approach to QCD thermodynamics which covers the broad region of strongly interacting matter properties from the hadron gas at low temperatures to a deconfined quark gluon plasma phase high temperatures.
Recently, finite temperature LQCD simulations have overcome the difficulties of reaching the low physical light quark masses and approaching the continuum limit which makes this theoretical laboratory now a benchmark for modelling QCD under extreme conditions \cite{Borsanyi:2010cj,Bazavov:2014pvz}.
In spite of this, effective models as such prove valuable for interpreting LQCD results as well as for extensions of QCD based ab-initio knowledge to the regions of the QCD phase diagram which are inaccessible to LQCD and high energy heavy ion collision experiments.
These are the regions high net baryon number and isospin densities at very low temperatures as probed by observations of astrophysical objects like compact stars as well as in neutron star mergers and supernovae collisions.

The aim of this paper is to construct a combined effective model reproducing the two known asymptotics of the finite temperature equation of state: the hadron resonance gas at low temperatures and the free quark-gluon plasma with perturbative corrections in accordance with the recent LQCD simulations.
This endeavour started about ten years ago with an ansatz for temperature- and mass dependent width in a hadron spectral function  \cite{Blaschke:2003ut,Blaschke:2005za}. 
As was then demonstrated in  \cite{Turko:2011gw} the nice description of LQCD data at that time stemmed from the fact that low-lying meson states below a mass threshold of $\sim 1$ GeV were not subject to spectral broadening so that at high-temperatures where the majority of states of the hadron resonance gas disappeared nevertheless a Stefan-Boltzmann limit was attained which accidentally coincided with the LQCD one because the number of low-lying hadronic degrees of freedom matches that of quarks and gluons. 
Note that the idea to use a spectral function approach to describe the dissociation of the hadrons in a resonance gas with increasing temperature has recently also been formulated in Ref.~\cite{Jakovac:2013iua}, but due to a lacking dynamical model of hadrons as bound states of quarks and gluons, the thermodynamics of the latter had there to be added by construction from a comparison with LQCD data. 

The next step \cite{Turko:2011gw} was to lift the lower mass threshold and let all hadrons be subject to spectral broadening leading to a decrease of their contribution to the total pressure and to add the pressure of the quark-gluon system according to the meanfield treatment of the Polyakov-loop generalized NJL model \cite{Ratti:2005jh}. 
Some of the arbitrariness of the choice of the temperature- and mass dependence of the spectral width $\Gamma(T,M)$ was removed in \cite{Turko:2014jta} where the latter was taken from a recently developed approach to chemical freeze-out \cite{Blaschke:2011ry}. 
This model could also make very nice predictions for the parton fractions in the temperature region of the crossover transition from hadronic to quark-gluon matter, in accordance with, e.g., Ref.~\cite{Nahrgang:2013xaa}. 
However, this model did still not yet provide a microscopic description of the temperature- and state- dependent spectral broadening of hadrons in the medium. 

In the present contribution we want to join these ideas for the description of a Mott-hadron resonance gas (MHRG) with a Beth-Uhlenbeck approach \cite{Beth:1937zz}
to pion and sigma meson dissociation \cite{Hufner:1994ma}
in its recent generalization from the NJL model \cite{Blaschke:2013zaa}
to the PNJL model case \cite{Wergieluk:2012gd,Dubinin:2013yga,Blaschke:2014zsa}.
The latter approach is based on the in-medium modification of (quark-antiquark-) scattering phase shifts
which encode the spectral properties of the complex-valued meson propagators.
For an alternative but equivalent formulation, see \cite{Yamazaki:2012ux}. 
In order to join the Beth-Uhlenbeck formulation of Mott dissociation with the hadron resonance gas, we will postulate a generic behaviour of the phase shifts in selected hadronic channels which are temperature dependent and embody the main consequence of chiral symmetry restoration in the quark sector: 
the lowering of the thresholds for the two- and three-quark scattering state continuous spectrum which triggers the transformation of hadronic bound states to resonances in the scattering continuum. 
The phase shift model is in accordance with the Levinson theorem (see, e.g., Refs. \cite{Dashen:1969ep, Zhuang:1994dw}) which results in the vanishing of hadronic contributions to the thermodynamics at high temperatures \cite{Wergieluk:2012gd}. 
While in this work we will restrict ourselves to a few states only, the model is of course applicable to the 
whole spectrum of hadronic states in the particle data book, which is presently under way \cite{Blaschke}.

\section{Quark and hadron mass spectrum}
The underlying quark and gluon thermodynamics is
described within a {$N_f = 2+1$}~PNJL model at the mean-field level
\begin{eqnarray}
\label{eq1}
P_{\rm PNJL}(T)=P_{\rm cond}(T)+P_{\rm FG}(T)+ {\cal U}(T),
\end{eqnarray}
with the $ N_f= 2+1 $ parameters determined analogously to 
\cite{Grigorian:2006qe}\footnote{The $m_s$ values in \cite{Grigorian:2006qe} are to be corrected, see 
{\tt http://3fcs.pendicular.net/psolver}} resulting in a dynamically generated vacuum light quark mass of $m(0) = 394.7$ MeV, and a strange quark mass of  $m_s(0)=604.745$~MeV for the parameters 
$\Lambda$ = 600 MeV, $G_s \Lambda^2=2.403$, $m_0= 5.4285$ MeV and $m_s = 136.459$~MeV. 
One obtains the temperature dependence of both the light quark mass $m(T)$ and the strange quark mass
$m_s(T)$ as a solution of the corresponding gap equations. 

We make the simplifying ansatz that the hadron masses are constant and temperature independent up to their Mott temperature, where they hit the two-quark threshold $m_{\rm thr,M}(T)$ (three-quark threshold $m_{\rm thr,B}(T)$) for mesons (baryons), respectively. After that temperature, we assume a linear rise of the resonance mass with temperature, being identified with the resonance width $ \Gamma_i(T)$, unique for all hadrons
\begin{eqnarray}\label{eq2}
	M_i(T) = M_i(0)+ \Gamma_i(T), \quad \Gamma_i(T) = (T-T_{\rm Mott,i})\theta(T-T_{\rm Mott,i}),	
\end{eqnarray}
where for the constant slope we choose $a = 2.5$.
\begin{figure}[h!]
\centerline{
\includegraphics[width=0.7\textwidth,angle=0]{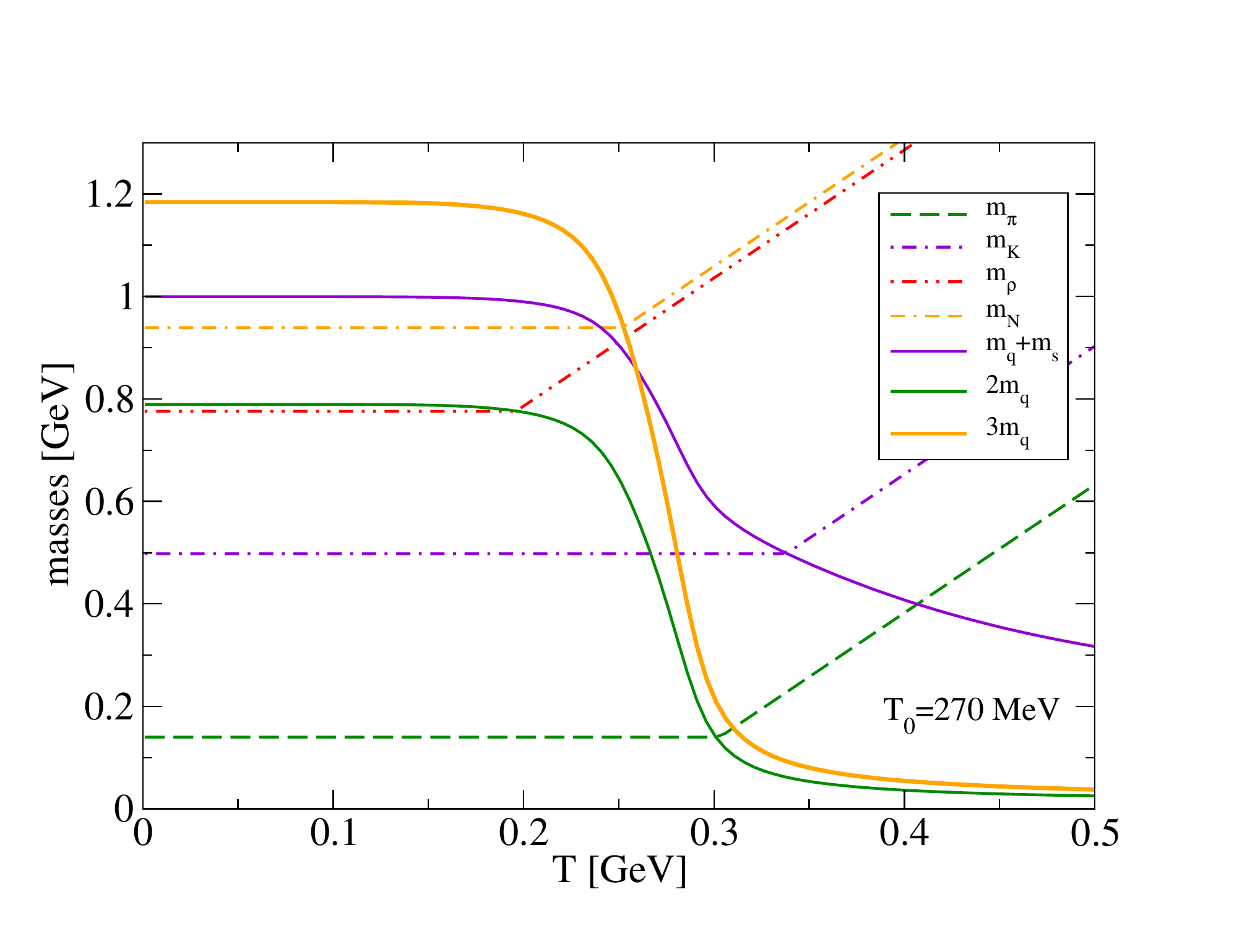}}
\caption{ Temperature dependent hadron mass spectrum used in the model.
\label{quarkmass}
}
\end{figure}
The Mott temperatures $T_{\rm Mott,i}$ can be determined as soon as the temperature dependence of the light quark mass $m(T)$ is known. They are determined from the
condition
\begin{eqnarray}\label{eq2}
	M_i(T_{\rm Mott,i}) &=& m_{\rm thr,i}(T_{\rm Mott,i}), \\
	m_{\rm thr,M}(T) &=& ( 2 - N_s)m(T)+N_s m_s(T),\\
	m_{\rm thr,B}(T) &=& ( 3 - N_s)m(T)+N_s m_s(T),
\end{eqnarray}
where $N_s= 0, 1, 2$ for mesons ($i = M$) and $Ns = 0,\ldots,3$ for baryons ($i = B$).
The  mass spectrum (\ref{eq2}) determines the corresponding Mott dissociation temperatures as  
$T_{\rm Mott,i} [{\rm MeV}]= 303, ~338,~196,~252$ (for hadrons $i = \pi, K, \rho, N$, resp.) after which
the hadrons became unbound state, see Fig.~\ref{quarkmass}.

\section{Mott effect and temperature dependent hadron phase shifts}	
We define temperature dependent phase shifts $\delta_i(s;T)$ for each hadronic channel (species) 
$i$ as functions of the Mandelstam variable $s = M^2$ by a generic behaviour which holds for both 
types of hadrons, mesons and baryons and encodes the following characteristic features
\begin{itemize}
		\item  a bound state with mass $M_i$ is recognized by a jump of the phase shift by $+\pi$ at $s=M_i^2$;
		\item  above the continuum threshold $s_{\rm thr,i} = m^2_{\rm thr,i}$ the phase shift drops to zero which is reached at $s_{\rm max,i} = s_{\rm thr,i} + N^2_i\Lambda^2$ where
$N_i =2$ for mesons and $N_i = 3$ for baryons;
		\item For $T>T_{\rm Mott,i}$ the step function of the true bound state gets smeared out over a width $\Gamma_i(T)$ resembling a resonance in the continuum.
	\end{itemize}
Our ansatz is given by  	
\begin{eqnarray}
\label{eq7}
\delta_{i}(s;T)&=&F(s) \left[  \frac{\pi}{2}+ \arctan\left(
 \frac{s-M_i^2(T)}{M_i(T)\Gamma_i(T)}\right) \right]
\Bigg\{
\theta ( m_{\rm thr,i}^2 - s) \nonumber \\
&+& \theta ( s - m_{\rm thr,i}^2 ) \theta (  m_{\rm thr,i}^2 + N_i^2\Lambda^2 - s )
\left[  \frac{ m_{\rm thr,i}^2 + N_i^2\Lambda^2 - s }{N_i^2\Lambda^2}    \right]  \Bigg\} ,
\end{eqnarray}
where the auxiliary function 
$F(s)=\sin(s/\Gamma^2) \Theta(\Gamma^2\pi/2 - s) +\Theta(s - \Gamma^2 \pi/2)$ 
has been introduced in order to ensure that the phase shift at $s=0$ shall always be zero, even at 
higher temperatures, where large values of the width parameter in the Breit-Wigner like ansatz 
would otherwise spoil this constraint. 
A microscopic calculation of the finite temperature pion phase shift which shows the correct behaviour has been performed in \cite{Blaschke:2013zaa}. It has been used to constrain the above ansatz for $F(s)$. 

\begin{figure}[!ht]
\centerline{
\includegraphics[width=0.7\textwidth,angle=0]{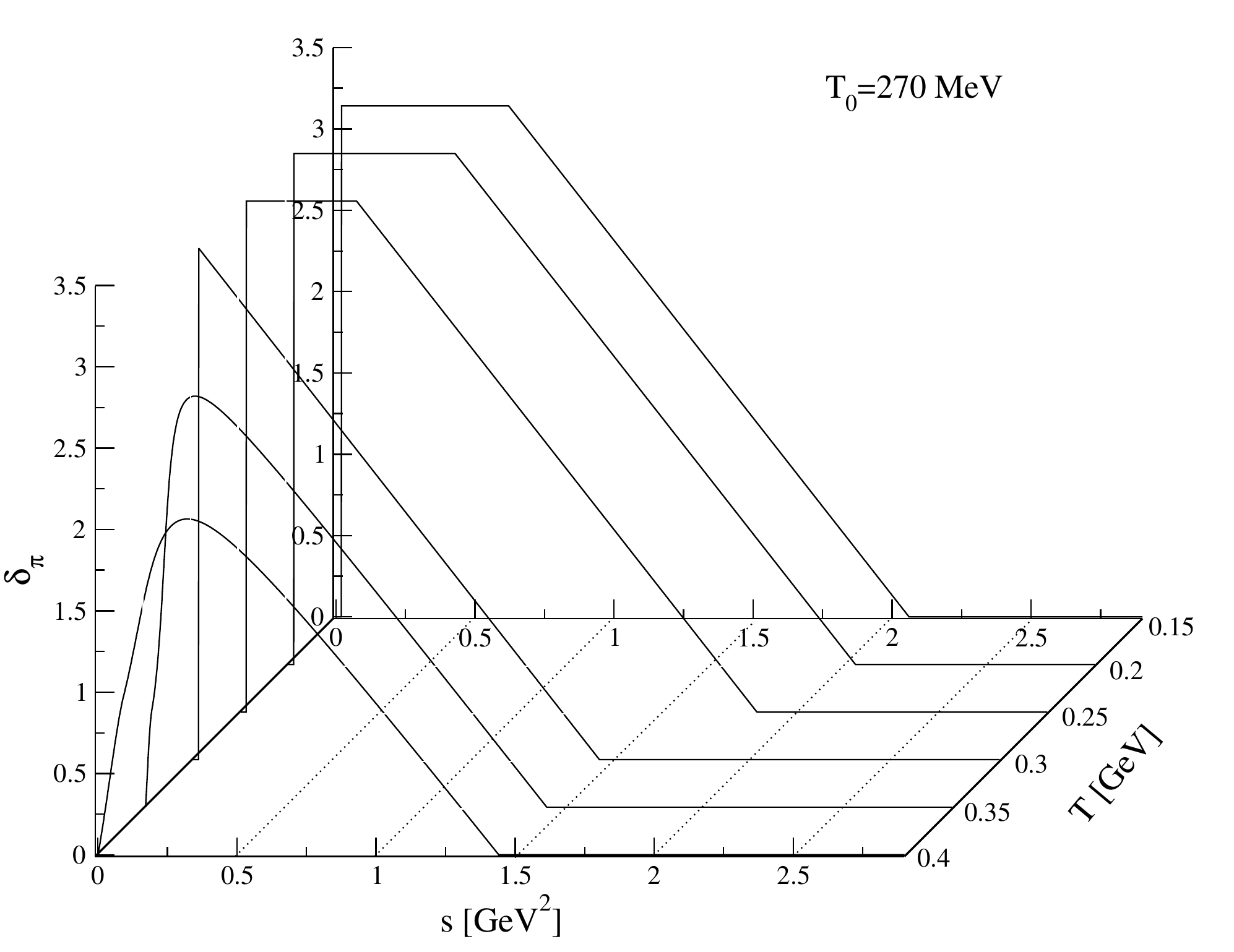}}
\caption{ Generic meson phase shift as defined by this model, here for the pion  ($N_i = 2, i = \pi$)
at six temperatures between 150 and 400 MeV.
\label{Ph}
}
\end{figure}

The behaviour of the phase shift (\ref{eq7}) as a function of the squared mass variable $s$ is shown for the example of the pion in Fig.~\ref{Ph}. The chosen values of six temperatures in steps of $50$ MeV, 
$T=150 \dots (50) \dots 400$ MeV cover the region of the Mott transition temperature $T_{\rm Mott,\pi}$ where the pion bound state (jump of $\delta_\pi(s)$ from zero to $\pi$ at $s=M_\pi^2$) merges with the edge of the continuum of scattering states at $s=m^2_{{\rm thr},\pi}$. 
At this temperature the mass gap (binding energy) vanishes which in Fig.~\ref{Ph} is recognizeable as 
the plateau at $\delta_{\pi}=\pi$ and consequently for higher temperatures $T>T_{\rm Mott,\pi}$ the 
pion phase shift starts at zero without reaching the value $\pi$ again. 
For higher temperatures the flattening of the hadron phase shift describes the vanishing of the hadronic correlation from the thermodynamics of the MHRG system.

\section{Mott hadron resonance gas (MHRG)}

For the thermodynamic potential, i.e. the pressure, of two-particle correlations in quark matter we have derived \cite{Blaschke:2013zaa} the formula
\begin{eqnarray}\label{eq8}	
	P_i(T) &=& d_i \int \frac{d^3p}{(2\pi)^3} \int^{\infty}_{0}
	\frac{d\omega}{\pi} f_i(\omega)\delta_i(\omega;T) \nonumber\\
  		&=& d_i\int^{\infty}_{0}\frac{dpp^2}{2\pi^2} \int^{\infty}_{0}\frac{ds}{2\pi}\frac{1}{\sqrt{p^2+s}}
f_i(\sqrt{p^2+s})\delta_i(s;T),
\end{eqnarray}
which we apply here to mesons and baryons, with $d_i$ being the degeneracy of the state $i$ and in the distribution function
%
$f_i(\omega) = 1/[\exp(\omega/T)\mp1]$,
where the upper (lower) sign holds for mesons (baryons). One can use $s = M^2$ and integrate over the mass variable $M$ instead of $s$, and changing the order of integrations to obtain
\begin{eqnarray}
P_i(T) &=& d_i\int^{\infty}_{0}\frac{dpp^2}{2\pi^2}
\int^{\infty}_{0}\frac{dM}{\pi}\frac{M}{\sqrt{p^2+M^2}}
f_i(\sqrt{p^2+M^2})\delta_i(M^2;T).
\label{eq:10}
\end{eqnarray}
By partial integration over $M$ we obtain
\begin{equation}\label{eq:10}
     P_i(T)  =  \mp d_i\int^{\infty}_{0}\frac{dpp^2}{2\pi^2}
\int^{\infty}_{0}d M\, T
\ln (1\mp e^{-\sqrt{p^2+M^2}/T})
\frac{1}{\pi}\frac{d\delta_i(M^2;T)}{dM}\ ,
\end{equation}
the generalized Beth-Uhlenbeck formula for the partial pressure of species $ i $ in the MHRG. 
When the phase shifts degenerate to a step function, i.e., when $m(T)\rightarrow \infty$, $\Gamma_i(T)\rightarrow 0$, so that $\delta_i(M^2;T) = \pi \theta(M -M_i)$, we have
\begin{eqnarray}
  \frac{1}{\pi}\frac{d\delta_i(M^2;T)}{dM} = \delta(M -M_i),
\end{eqnarray}
so that the $M$-integration becomes trivial and gives
\begin{eqnarray}
P_i(T) = \mp d_i \int^{\infty}_{0}\frac{dp\, p^2}{2\pi^2}\,
T \ln\Bigg( 1 \mp e^{-\sqrt{p^2+M^2_i}/T} \Bigg),
\label{eq:12}
\end{eqnarray}
the relativistic ideal quantum gas for a particle species $i$ with mass $M_i$.
For the total pressure of the model, we have 
%
\begin{eqnarray}
P(T) = \sum_{i = M,B}{P_i(T)}+P_{\rm PNJL}(T) +P_{\rm pert} (T)~,
\label{eq:15}
\end{eqnarray}
where the pressure of the quarks and gluons (\ref{eq1}), and a perturbative pressure correction 
have been added. 
The latter results from ${\cal O}(\alpha_s)$ interactions of quarks and gluons in a plasma for momenta exceeding the cutoff $\Lambda$ to which the nonperturbative interactions in the PNJL model are restricted. 
Details are described in Refs.~\cite{Turko:2011gw,Turko:2014jta}.
Note that at variance with these references, we use here a temperature dependent, regularized 
running coupling \cite{Blaschke:2005jg,Shirkov:2002td}
\begin{eqnarray}
\label{eq16}
\alpha(T) = \frac{12\pi}{11N_c-2N_f}\left( \frac{1}{\ln(r^2/c^2)} - \frac{c^2}{r^2-c^2} \right) ,
\end{eqnarray}
where $r =3.2 T$, $c = 350$ MeV and $N_c=N_f=3$.

\begin{figure}[h!]
\centerline{
\includegraphics[width=0.49\textwidth,angle=0]{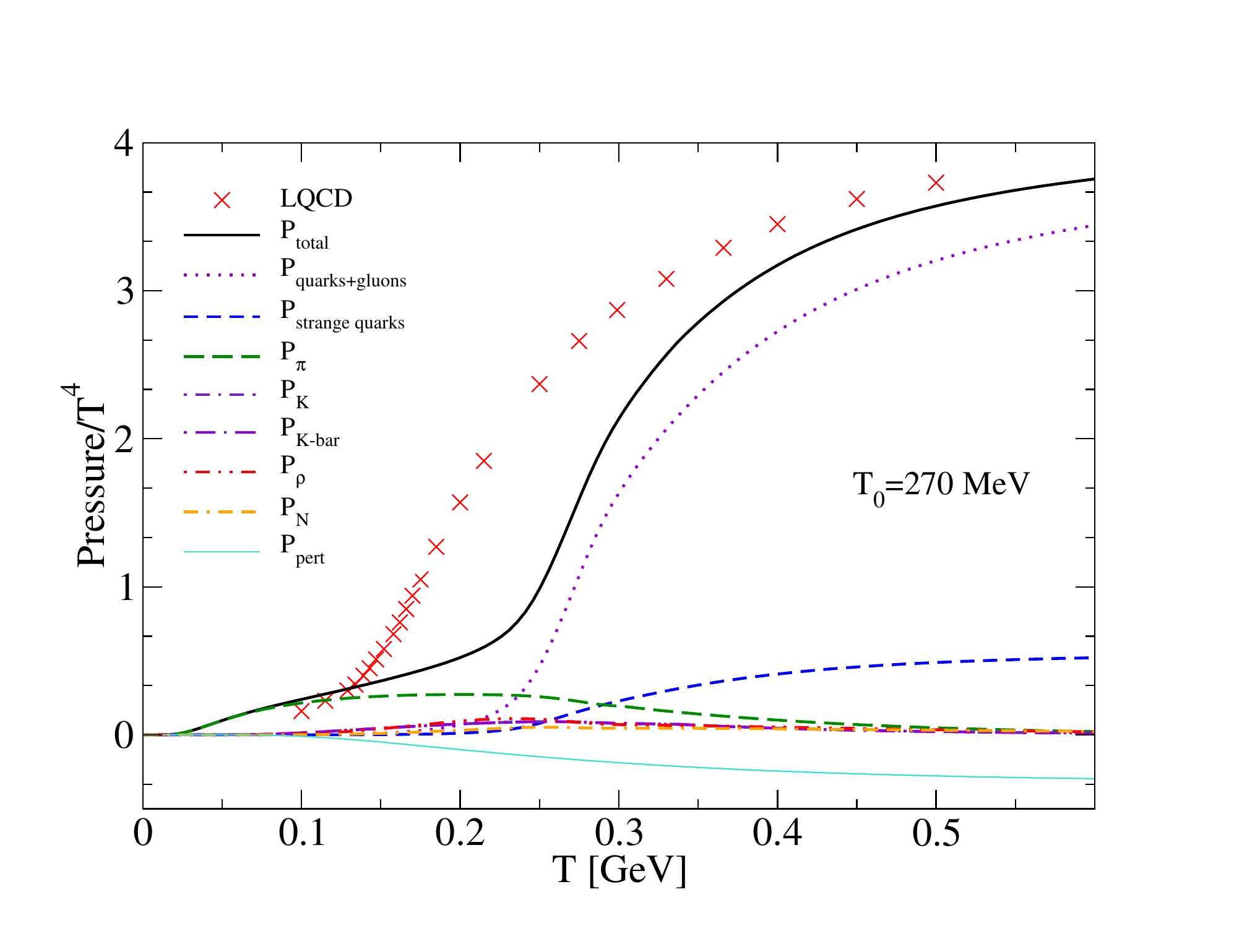}
\includegraphics[width=0.51\textwidth,angle=0]{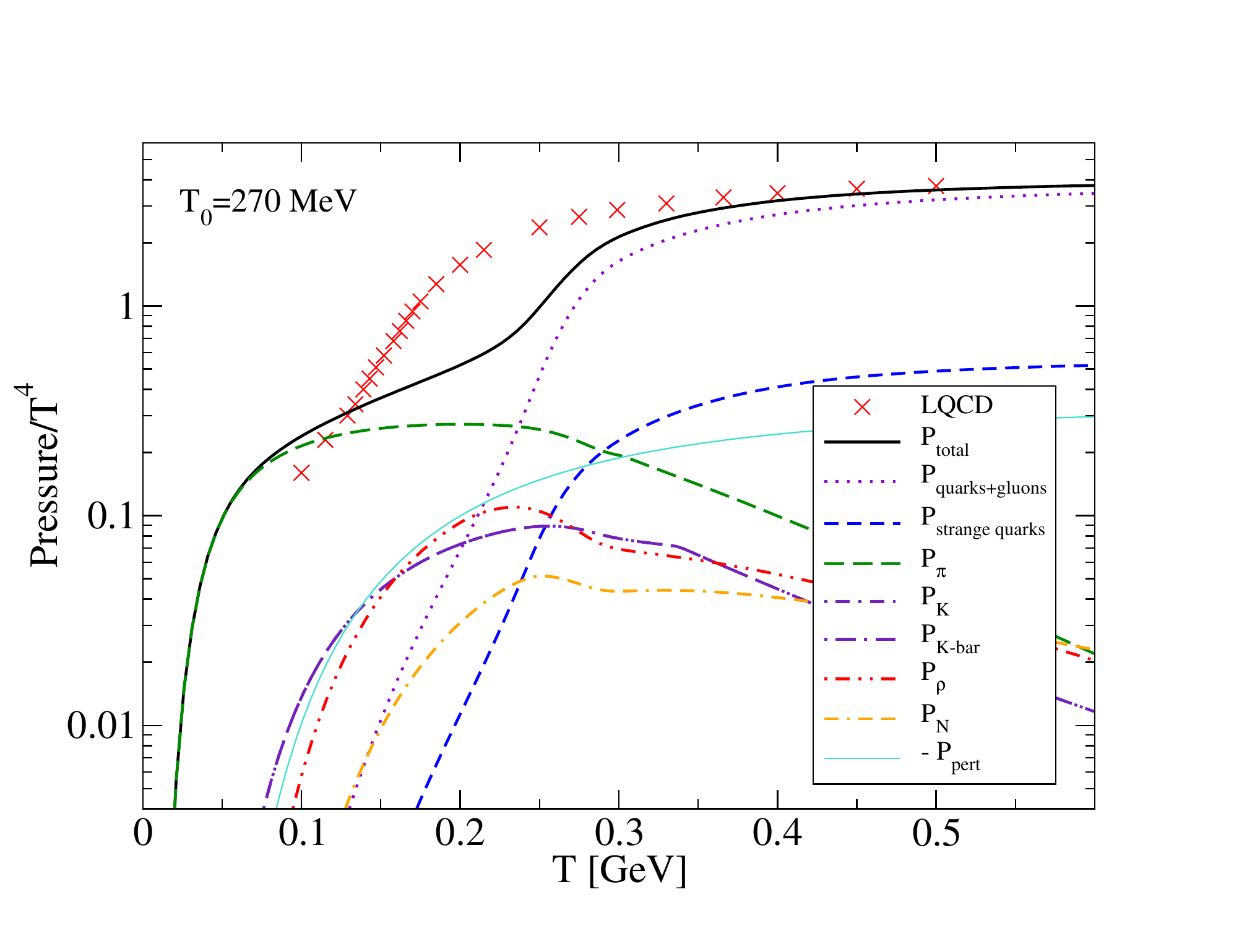}}
\caption{Temperature dependence of the pressure for linear scale (left panel) and
 logarithmic scale (right panel) \cite{Borsanyi:2010cj}.\label{pressure}
}
\end{figure}

In Fig.~\ref{pressure} we compare the total pressure of the MHRG model (black solid line) with LQCD data from Ref.~\cite{Borsanyi:2010cj} (star symbols) and show also the partial pressure contributions from all components in the linear (left panel) and logarithmic scale (right panel).
With the present schematic model, we can quantify both asymptotic limits, the pion gas limit at low temperatures $T < 100$ MeV and the quark-gluon plasma limit at high temperatures $T>300$ MeV.
In the region between these temperatures we observe a discrepancy between the model and LQCD data which is attributed to mainly two missing aspects: i) the inclusion of higher lying hadronic resonances and ii) the back reaction of the MHRG on the temperature dependence of the quark masses and thus on the continuum thresholds which trigger the onset of the Mott dissociation of hadrons. 

\section{Conclusions}			
				
In this work an effective model is constructed which is capable of reproducing basic physical characteristics of the hadron resonance gas at low temperatures and embody the crucial effect of hadron dissociation by the Mott effect.
The generalized Beth-Uhlenbeck form of the partial pressures is constructed for each hadronic channel. Numerical results show that the simplifying ansatz for the temperature dependence of both, the mass spectrum and the phase shifts of hadronic channels give results in qualitative agreement with recent ones from LQCD \cite{Borsanyi:2010cj}.
The differences at the present level of sophistication of our effective model are well understood and basically two conclusions for its further development can be drawn.

First, one can include higher lying hadronic states into the model.
Second, and most importantly, the back reaction of the hadron resonance gas on the chiral condensate and thus to the quark masses and the continuum thresholds which derive from it must be taken into account.
This latter point will lower the onset of chiral restoration and hadron dissociation towards the result from
lattice QCD. 
It shall further pronounce the character of a smooth crossover transition between MHRG and QGP by quantifying the relative contributions of hadronic and partonic degrees of freedom (partial fractions) to the QCD thermodynamics.
	
\subsection{Acknowledgements}

We acknowledge support from the Polish National Science Center (NCN) under contract
No. UMO-2011/02/A/ST2/00306 (D.B., A.D.) and No. DEC-2013/10/A/ST2/00106 (L.T.).
The work of D.B. was supported in part by the Polish Ministry of Science and Higher Education under
grant No. 1009/S/IFT/14 and A.D. is grateful for support from the University of Wroclaw under internal
grant No. 2470/M/IFT/14.


\begin{thebibliography}{99}
	
\bibitem{Borsanyi:2010cj} 
  S.~Bors\'{a}nyi, G.~Endr\"odi, Z.~Fodor, A.~Jakov\'{a}c, S.~D.~Katz, S.~Krieg, C.~Ratti and K.~K.~Szab\'{o},
  {\it The QCD equation of state with dynamical quarks.}
  JHEP (2010) {\bf 1011}, 077.
	
\bibitem{Bazavov:2014pvz} 
  A.~Bazavov {\it et al.}  [HotQCD Collaboration],
  {\it Equation of state in (2+1)-flavor QCD.}
  Physical Review D  (2014) {\bf 90}, 094503.
	
\bibitem{Blaschke:2003ut}
  D.~B.~Blaschke and K.~A.~Bugaev,
  {\it Hadronic correlations above the chiral /
  deconfinement transition.}
  Fizika B (2004) {\bf 13}, 491.
  
\bibitem{Blaschke:2005za}
  D.~B.~Blaschke and K.~A.~Bugaev,
 {\it Thermodynamics of resonances with finite width.}
  Physics of Particles and Nuclei Letters  {\bf 2} (2005) 305.

\bibitem{Turko:2011gw} 
  L.~Turko, D.~Blaschke, D.~Prorok and J.~Berdermann,
  {\it Mott-Hagedorn Resonance Gas and Lattice QCD Results.}
  Acta Physica Polonica Supplement  (2012) {\bf 5}, 485.

\bibitem{Jakovac:2013iua} 
  A.~Jakov\'{a}c,
  {\it Hadron melting and QCD thermodynamics.}
  Physical Review D  (2013) {\bf 88}, 065012.

\bibitem{Ratti:2005jh}
  C.~Ratti, M.~A.~Thaler and W.~Weise,
  {\it Phases of QCD: Lattice thermodynamics and a field theoretical model.}
  Physical Review D (2006) {\bf 73}, 014019.

\bibitem{Turko:2014jta} 
  L.~Turko, D.~Blaschke, D.~Prorok and J.~Berdermann,
  {\it Effective degrees of freedom in QCD thermodynamics.}
  EPJ Web Conferences (2014) {\bf 71}, 00134.

\bibitem{Blaschke:2011ry}
  D.~B.~Blaschke, J.~Berdermann, J.~Cleymans and K.~Redlich, 
  {\it Chiral condensate and chemical freeze-out.}
  Physics of Particles and Nuclei Letters (2011) {\bf 8}, 811.

\bibitem{Nahrgang:2013xaa} 
  M.~Nahrgang, J.~Aichelin, P.~B.~Gossiaux and K.~Werner,
  {\it Influence of hadronic bound states above $T_c$ on heavy-quark observables in Pb + Pb collisions at at the CERN Large Hadron Collider.}
  Physical Review C  (2014) {\bf 89}, 014905.

\bibitem{Beth:1937zz} 
  E.~Beth and G.~Uhlenbeck,
  {\it The quantum theory of the non-ideal gas. II. Behaviour at low temperatures.}
  Physica (1937) {\bf 4}, 915.

\bibitem{Hufner:1994ma} 
  J.~H\"ufner, S.~P.~Klevansky, P.~Zhuang and H.~Voss,
  {\it Thermodynamics of a quark plasma beyond the mean field: A generalized Beth-Uhlenbeck approach.}
  Annals of Physics (1994) {\bf 234}, 225.

\bibitem{Blaschke:2013zaa}
  D.~Blaschke, D.~Zablocki, M.~Buballa, A.~Dubinin and G.~R\"opke,
  {\it Generalized Beth--Uhlenbeck approach to mesons and diquarks in hot, dense quark matter.}
  Annals of Physics (2014)   {\bf 348}, 228.

\bibitem{Wergieluk:2012gd} 
  A.~Wergieluk, D.~Blaschke, Y.~L.~Kalinovsky and A.~Friesen,
  {\it Pion dissociation and Levinson`s theorem in hot PNJL quark matter.}
  Physics of Particles and Nuclei Letters (2013) {\bf 10}, 660.

\bibitem{Dubinin:2013yga}
  A.~Dubinin, D.~Blaschke and Y.~L.~Kalinovsky,
  {\it Pion and sigma meson dissociation in a modified NJL model at finite temperature.}
  Acta Physica Polonica B Supplement  (2014) {\bf 7},  215.

\bibitem{Blaschke:2014zsa} 
  D.~Blaschke, A.~Dubinin and M.~Buballa,
  {\it Polyakov-loop suppression of colored states in a quark-meson-diquark plasma.}
  [{\tt arXiv:1412.1040 [hep-ph]}].

\bibitem{Yamazaki:2012ux} 
  K.~Yamazaki and T.~Matsui,
  {\it Quark-Hadron Phase Transition in the PNJL model for interacting quarks.}
   Nuclear Physics A (2013) {\bf 913}, 19.

\bibitem{Dashen:1969ep} 
  R.~Dashen, S.~K.~Ma and H.~J.~Bernstein,
  {\it S Matrix formulation of statistical mechanics.}
  Physical Review (1969) {\bf 187}, 345.

\bibitem{Zhuang:1994dw} 
  P.~Zhuang, J.~H\"ufner and S.~P.~Klevansky,
  {\it Thermodynamics of a quark - meson plasma in the Nambu-Jona-Lasinio model.}
  Nuclear Physics A (1994) {\bf 576}, 525.

\bibitem{Blaschke}
D.~Blaschke, A.~Dubinin and L.~Turko, in preparation.

\bibitem{Grigorian:2006qe}
  H.~Grigorian,
  {\it Parametrization of a nonlocal, chiral quark model in the instantaneous three-flavor case: Basic formulas and tables.}
  Physics of Particles and Nuclei Letters  (2007) {\bf 4}, 223.


\bibitem{Blaschke:2005jg} 
  D.~Blaschke, O.~Kaczmarek, E.~Laermann and V.~Yudichev,
  {\it Heavy quark potential and quarkonia dissociation rates.}
  European Physical Journal C (2005) {\bf 43}, 81.

\bibitem{Shirkov:2002td} 
  D.~V.~Shirkov,
 {\it On the Fourier transformation of renormalization invariant coupling.}
  Theoretical and Mathematical Physics (2003) {\bf 136}, 893.

\end{thebibliography}
\end{document}